\documentstyle[12pt,a4,graphicx]{article}

\newcommand{\ee}{\end{equation}}
\newcommand{\be}{\begin{equation}}

\newcommand{\N}{{\rm I \hspace{-1mm} N}}
\newcommand{\Z}{{\rm Z \hspace{-1.2mm} Z}}
\newcommand{\R}{{\rm I \hspace{-0.9mm} R}}
\newcommand{\CE}{{\cal E}}
\newcommand{\CL}{{\cal L}}
\newcommand{\CH}{{\cal H}}
\newcommand{\CN}{{\cal N}}
\newcommand{\isomo}{=}
\newcommand{\bphi}{\bar{\varphi}}
\newcommand{\bA}{\bar{A}}
\newcommand{\lPhi}{\bar{\Phi}}
\newcommand{\lA}{\bar{A}}
\newcommand{\tphi}{\check{\varphi}}
\newcommand{\tA}{\check{A}}
\newcommand{\aphi}{\tilde{\varphi}}
\newcommand{\aPhi}{\tilde{\Phi}}
\newcommand{\aA}{\tilde{A}}
\newcommand{\sgn}{\mbox{sgn}}
\newcommand{\tr}{\mbox{tr}}
\newcommand{\Ncs}{{\cal N}_{CS}}

\begin{document}

\thispagestyle{empty}
\vspace*{0.5cm}
\begin{center}
{\LARGE\bf $d$ dimensional $SO(d)$--Higgs Models with 
Instanton and Sphaleron: $d=2,3$ } 
\\[19mm] {\large
B. Kleihaus\raisebox{0.8ex}{\small a,}\footnote{E--mail:
kleihaus@stokes1.thphys.may.ie},
D.H. Tchrakian\raisebox{0.8ex}{\small a,b,}\footnote{E--mail:
tigran@thphys.may.ie} and
F. Zimmerschied\raisebox{0.8ex}{\small a,}\footnote{E--mail:
zimmers@thphys.may.ie}}
\\[1cm]
\raisebox{0.8ex}{\small a} {\it Department of Mathematical Physics \\
National University of Ireland, Maynooth, Ireland}
\\[4mm] \raisebox{0.8ex}{\small b}
{\it School of Theoretical Physics, Dublin Institute for Advanced studies, 
\\ 10 Burlington Road, Dublin 4, Ireland}
\vfill
{\bf Abstract}
\end{center}

\noindent
The Abelian Higgs model and the Georgi-Glashow model in 2 and 3 
Euclidean dimensions respectively,
support both finite size instantons and sphalerons. The instantons 
are the familiar Nielsen--Oleson
vortices and the 't~Hooft--Polyakov monopole solutions respectively. 
We have constructed the
sphaleron solutions and calculated the Chern-Simons charges $\Ncs$ 
for sphalerons of both models
and have constructed two types of noncontractible loops 
between
topologically distinct vacuua. In the 3 dimensional model, the 
sphaleron and the vacuua have zero magnetic and electric flux
while the configurations on the loops have non vanishing magnetic flux.
\\[2cm]
\newpage

\section{Introduction}

In the semiclassical treatment of quantum field theories, 
instantons \cite{BPST,tH,dilute}
play the important role of providing the tunneling between 
topologically inequivalent
vacuua in an essentially nonperturbative framework. Instantons 
are classical solutions
to the Euler-Lagrange equations with Euclidean time, whence comes 
the tunneling
interpretation. Another, closely related mechanism
for vacuum to vacuum transitions but now at non--zero temperature, is 
provided by the
(Euclidean) static solutions of the field equations called 
sphalerons \cite{Manton}. In
contrast to instantons, sphalerons are unstable solutions and 
provide a classical
rather than quantum (tunneling) transition over the energy barrier 
separating the two
vaccua. In this context it is very useful to consider another class of 
classical
solutions, which are periodic in (Euclidean) time, with the period being
identified with the inverse of the temperature. These are the periodic 
instantons
as defined in Ref.~\cite{KRT}. Thus at zero temperature the period of 
the periodic
instanton becomes infinite, rendering it non-periodic, which can be 
identified as the
instanton itself.

Now a given field theoretic model supporting sphalerons 
\cite{Manton,Klinkhamer} and 
hence also
periodic instantons \cite{KRT}, may or may not support (zero--temperature or 
infinite period) instanton. In
the case where a finite--size instanton exists, it can 
be arrived at as the period of
the periodic instanton tends to infinity. In the case however 
where no finite size
instanton exists, the situation is more complicated and the 
so--called constrained
instantons \cite{A} must be employed. The consequences of a 
theory being of one type
or the other are thought to be potentially important. 

There has been some exploratory work done in this direction, 
in the context of the $1+1$
dimensional scale--breaking $O(3)$ sigma model \cite{MW} which 
does not support finite
size instantons, and its skyrmed version \cite{PMTZ} which 
does so. The study of the
periodic instantons in these two models from this viewpoint 
was carried out in refs.~\cite{HMT,KT} respectively.

In view of the above, it is interesting to construct and 
study models that can support
both {\it instantons} and {\it sphalerons}, as a first 
step before studying the {\it periodic
instantons} interpolating them. Our aim in this paper 
is to do just this, for a class of
$d$-dimensional $SO(d)$ Higgs models in which the 
Higgs field is a $d$-component vector of
the $SO(d)$, for the two cases of $d=2$ and $d=3$. 
What distinguishes such models is that
their instantons result in curvature field strengths 
that exhibit {\it inverse square}
behaviour. This property contrasts with the pure-gauge 
behaviour of the arbitrary--scale
Yang-Mills (YM) instantons \cite{BPST} and can result 
in far reaching (physical) consequences.
In the $d=3$ case, it leads to a dilute Coulomb gas of 
instantons, as shown by Polyakov~\cite{dilute}
long ago, while the corresponding instantons in $d=4$ 
share this property \cite{OT} and can
possibly also enable the construction of a dilute 
Coulomb gas \cite{so4inst}. This is our physical
justification for making a systematic study of these 
Higgs models, and we concentrate on the
$d=2,3$ cases here. Thus both models under consideration 
here are the familiar ones, namely
the Abelian Higgs model in $d=2$ and the Georgi--Glashow 
model in $d=3$.

The instantons in these two, $1+1$ and $2+1$ dimensional 
models, are the well
known topologically stable Nielsen-Olesen vortices \cite{NO} 
and the 't Hooft--Polyakov
monopoles \cite{tHP}, respectively. It is therefore the 
sphaleron solutions to these models that
are the remaining entities to be studied. In this connection, 
it is true that the sphalerons
of the Abelian Higgs model were studied extensively long 
ago \cite{BS,C}, but we repeat it here for the sake of completeness so
that both cases $d=2$ and $d=3$ be treated similarly,
and, because the presentation of the results in the literature can be refined.
In particular we clarify the situation
with respect to the question of periodic boundary conditions in the
spacelike coordinate used in the literature~\cite{BS,C}.
The sphaleron in the $d=3$ case has been studied recently~\cite{Copeland}.
We have carried out the analysis of the sphaleron solutions in both models
employing both the non-contractible loop (NCL) used by Manton~\cite{Manton} for
the Weinberg-Salam model, to which we refer in this paper as the
{\it geometric} loop construction, as well as the the finite energy path
method of Akiba et al~\cite{AKY}, to which we refer as {\it boundary} loop
construction.

In the study of the sphaleron solutions, a central role is played
by the Chern--Simons number, which in the Weinberg--Salam model is calculated
from the second Chern--Pontryagin (CP) density. In  $d$ dimensional
Higgs models we consider, the natural
candidate for the latter is the dimensionally reduced CP density
of the YM field on $\R^d \times S^{4p-d}$, where $4p>d$~\cite{desc}. 
The $d$ dimensional
model is decided by $p$ such that the action density is bounded from below
by dimensionally reduced the CP density in question. 
The simplest such model corresponds to
the case where $4p$ is the smallest number greater than $d$. Both the Abelian
Higgs model and the Georgi--Glashow models considered in this work are the
simplest models whose CP densities are arrived at by the corresponding
dimensional descent of the {\it second} CP density of the $SU(2)$ YM field.

The analysis of the Abelian Higgs system in 2
dimensions is presented in Section 2. In
Subsection 2.1 and 2.3 respectively, the {\it geometric} and {\it boundary}
loop constructions are presented, while in Subsection 2.2 a discussion of our
constructions is contrasted with what is usually given in the literature, and
the differences are commented on.

The analysis of the Georgi-Glashow model in 3 dimensions is presented in
Section 3. In Subsection 3.1 the {\it boundary} loop construction is presented.
The {\it geometric} loop construction is presented in Subsection 3.2 . Since
the sphaleron solution of this model turns out to be effectively the solution
of an Abelian Higgs model, a family of arbitrary vorticity $N$ sphalerons are
contructed in Subsection 3.2 . Subsection 3.3 is devoted to a brief discussion
of the magnetic and electric properties of this sphaleron and the
configurations on the geometric loop. In section 4, we
give a summary of our results.

\section{$d=2$: Sphalerons in the Abelian Higgs model}

Our objective in this section is to construct noncontractible loops (NCL)
between two topologically neighbouring vacua which feature the sphaleron at 
the top of the energy barrier. Our constructions run exactly parallel to those
of Manton \cite{Manton} and Akiba {\em et.al} \cite{AKY} respectively.
This differs from the analysis of Bochkarev and Shaposhnikov \cite{BS} in
that the latter employ periodicity in the space variable to construct the
NCL.

The $SO(2)$ Abelian Higgs model in $d=2$ spacetime dimensions is
given by the Euclidean Lagrange density 
\begin{equation}
\CL=\frac14F_{\mu\nu}^2+\frac12|D_{\mu}\varphi|^2+
\frac{\lambda_0}{32}(1-|\varphi|^2)^2
\label{a1}
\end{equation} 
with $\varphi=\phi^1+i\phi^2$, 
$D_{\mu}\varphi=\partial_{\mu}\varphi+iA_{\mu}\varphi$, i.e.\ we used the 
identity $SO(2)=U(1)$ to write the Higgs dublett $\phi^i$ as one complex
field. 
 
Adding the inequalities
\begin{equation}
\left(F_{\mu\nu}-\frac{\sqrt{\lambda_0}}{4}
\epsilon_{\mu\nu}(1-|\varphi|^2)\right)^2\ge 0, \qquad  
\left|D_{\mu}\varphi-i\epsilon_{\mu\nu}D_{\nu}\varphi\right|^2\ge 0,
\label{ac1}
\end{equation}
the cross terms of the squares yield a lower bound for the 
Euclidean action, $\CL\ge \varrho$ which becomes a topological lower bound,
i.e.\ $\varrho$ can be written as total divergence
\begin{equation}
\varrho = \partial_{\mu}\Omega_{\mu}, \quad
\Omega_{\mu}=\frac14\epsilon_{\mu\nu}\left(A_{\nu}+
i\varphi^*D_{\mu}\varphi\right)
\label{a3}
\end{equation}
if $\lambda_0=1$.
$\Omega_{\mu}$ is the Chern--Simons form of this model. 
It consists of a gauge 
dependent and a gauge independent part which is a typical feature of 
Higgs theories in even spacetime dimensions \cite{desc}. 

The topological lower bound (which can be generalised to $\lambda_0\neq 1$)
ensures the stability of the instantons of the model, the Abelian Higgs
vortex, which can be found using the radially symmetric ansatz
\begin{equation}
\varphi=h(r)e^{-iN\theta} ,\qquad A_{\mu}=\frac{a(r)-N}{r} \ , 
\epsilon_{\mu\nu}\hat{x}_{\nu}
\label{aa1}
\end{equation}
with $r^2 = x^2_0 +x^2_1=t^2+x^2$,
and integrating the Euler--Lagrange equations of the resulting radial
subsystem Lagrangian
\begin{equation}
L_0=\pi \left\{   {1\over r}a'^2 +r\left(h'^2 + \frac{a^2 h^2}{r^2}
\right) +\frac{\lambda_0}{16} (1-h^2)^2    \right\} \ .
\label{abhiggsvortex}
\end{equation}
The corresponding instanton solutions are self--dual provided $\lambda_0=1$,
then they saturate the 
inequality $\CL\ge\partial_{\mu}\Omega_{\mu}$ 
and hence are characterised by integer Chern--Simons number (also
called Chern--Pontryagin charge in the context of instanton physics)
\begin{equation}
\Ncs= \frac{2}{\pi}\int_{-\infty}^{\infty}dt\int_{-\infty}^{\infty}dx\:
\partial_{\mu}\Omega_{\mu}=
\frac{2}{\pi}\int_{\R^2}\partial_{\mu}\Omega_{\mu}d^2x=
\frac{2}{\pi}\lim_{r\rightarrow\infty}\int_{S^1}\Omega_{\mu}dS_{\mu}.
\label{aa3}
\end{equation}
This can be arrived at by subjecting the second Chern-Pontryagin class of the
$SU(2)$ YM field on $\R^2 \times S^2$ to dimensional descent~\cite{desc}.

The Chern--Simons number depends only 
on the behaviour of the instanton at infinity, reflecting the topological
properties of the mapping 
\begin{equation}
\varphi_{inst}^{\infty}:S^1_{spacetime}\rightarrow S^1_{Higgs}.
\label{ab1}
\end{equation}

In contrast to the instanton, the sphaleron is a {\em ``static''} 
object, i.e.\ it does not depend on the Euclidean time. Nontheless, it is an 
{\em Euclidean} spacetime 
object closely related to the instanton as it is the extremum 
of the energy functional of a set of 
static configurations connecting distinct vacua in a topologically nontrivial 
way, i.e.\ a {\em noncontractible loop} (NCL). 
The NCL, parametrised in 
terms of Euclidean time, is an instanton--like object 
and the loop as a whole having integer 
topological number. 
The Chern-Simons number $\Ncs$ is also parametrized in terms of the Euclidean
time, 
\begin{equation}
\Ncs(t_0)= \frac{2}{\pi}\int_{-\infty}^{t_0}dt\int_{-\infty}^{\infty}dx\:
\partial_{\mu}\Omega_{\mu},
\label{Ncstime}
\end{equation}
hence the loop can be parametrizes in terms of $\Ncs$, the latter taking 
on all values between $0$ and $1$.

To construct and study the sphaleron, we start from the static energy which in
the temporal gauge ($A_0=0$, $A:=A_1$) is given by
\begin{equation}
\CE[\varphi,A]=\int\left[\frac12|(\partial_x+iA)\varphi|^2+
\frac{1}{32}(1-|\varphi|^2)^2\right]dx \ .
\label{a4}
\end{equation}
Choosing the particular ansatz $\varphi=\phi\in\R$, $A=0$, this 
reduces to the well--known real $\varphi^4$ energy functional 
\begin{equation}
\CE_{sph}[\phi]=\int\left[\frac12(\phi')^2
+\frac{1}{32}(1-\phi^2)^2\right]dx
\label{aaa4}
\end{equation} 
which is minimized by the kink/antikink solutions 
\begin{equation}
\phi_{\pm}(x)=\pm\tanh\left(\frac{x-x_0}{4}\right).
\label{ad1}
\end{equation}
The $\phi^4$ kink on its own is a stable soliton, but it becomes instable 
as soon as it is embedded in complex isospace which can be shown explicitly 
by investigating the fluctuation spectrum \cite{FH}.

\subsection{The geometrical loop construction}

To show that 
$(\varphi,A)_{sph}=(\phi_{\pm},0)$ are sphaleron configurations 
of the  Abelian Higgs model, we construct the corresponding NCL
of finite energy configurations connecting two vacua through the 
sphaleron in a  topologically nontrivial way.

{\em Vacua} are static configurations with zero energy, i.e.\ they are given by
$\varphi=g$, $A=ig^{-1}\partial_xg$ with $g\in U(1)$. 
We use the remaining gauge 
freedom of the theory to fix the vacuum to $(\varphi,A)_{vac}=(1,0)$ and 
comment on the choice of gauge later on.

Following the geometrical loop construction of Manton \cite{Manton}, we 
consider first the topological properties of the Higgs field at infinity.
One dimensional space at infinity shrinks to the discrete set 
$(\R)^{\infty}\isomo S^0_{space}\isomo\{\pm1\}$. Hence one has to distinguish 
two mappings $\varphi^{+\infty}$, $\varphi^{-\infty}$ instead of one mapping 
$\varphi^{\infty}$ which depends on continous angular coordinates in higher 
space dimensions. To construct a geometrical NCL \cite{Manton}, 
we introduce a single loop 
parameter $\tau\in S^1_{loop}$ as additional degree of freedom such that 
\begin{equation}
\varphi^{+\infty}:S^1_{loop} \rightarrow S^1_{Higgs} 
\label{ae1}
\end{equation}
is a topologically 
nontrivial mapping which we choose to be 
\begin{equation}
\varphi^{+\infty}:\tau\mapsto e^{2i\tau},
\label{af1}
 \end{equation}
whereas $\varphi^{-\infty}\equiv1$ is chosen to be topologically trivial.
The finite energy condition then also fixes the gauge field at infinity as 
the covariant space derivative in (\ref{a4}) has to vanish, hence 
\begin{equation}
A^{\pm\infty}:=
A(r\rightarrow\pm\infty)
=-i(\varphi^{\pm\infty})^{-1}\partial_x\varphi^{\pm\infty}=0. 
\label{ag1}
\end{equation}

The general Manton loop ansatz constructed using the topological ingredients 
$\varphi^{+\infty}$, $A^{\pm\infty}$ is now given by 
\begin{equation}
\bphi=(1-h(x))\psi+h(x)\varphi^{+\infty},\quad A=fA^{\pm\infty}=0 
\label{ah1}
\end{equation}
where
$\psi$ has to be chosen such that the loop starts and ends in the vacuum and 
reaches the sphaleron for $\tau=\frac{\pi}{2}$, hence $\bphi|_{\tau=0}=
\bphi|_{\tau=\pi}=1$,
$\bphi|_{\tau=\frac{\pi}{2}}=h\varphi^{+\infty}|_{\tau=\frac{\pi}{2}}=-h$.
In the $SO(2)$ model, 
\begin{equation}
\psi=\cos^2\tau+i\sin\tau\cos\tau
\label{ai1}
\end{equation}
is a convenient choice. Moreover,
$\bphi(x\rightarrow\pm\infty)=\varphi^{\pm\infty}$ requires
$h(x\rightarrow\pm\infty)=\pm 1$. 
The loop is then given by
\begin{equation}
\bphi(\tau,x)=e^{i\tau}[\cos\tau+ih(x)\sin\tau],\qquad \bA\equiv 0.
\label{a5}
\end{equation}
Inserting the ansatz (\ref{a5}) into the static energy functional (\ref{a4})
yields
\begin{equation}
\CE[\bphi,\bA]=\bar{\CE}_{\tau}[h]=\sin^2\tau \int\left[\frac12(h')^2+
\frac{1}{32}\sin^2\tau(1-h^2)^2\right]dx
\label{a6}
\end{equation}
which for $\tau=\frac{\pi}{2}$ reduces to the $\varphi^4$ model,
$\bar{\CE}_{\tau=\frac{\pi}{2}}=\CE_{sphal}=\frac{1}{3}$. 

For any fixed value of $\tau\in[0,\pi]$, $\bar{\CE}_{\tau}[h]$ is 
{\em minimised} by 
\begin{equation}
h_{\tau}=\tanh\left(\sin\tau\frac{x-x_0}{4}\right).
\label{aa6}
\end{equation} 
The energy along the resulting minimal energy loop is
\begin{equation}
\bar{\CE}(\tau)=\frac13\sin^3\tau 
\label{ab6}
\end{equation}
which has a {\em maximum} for
$\frac{\pi}{2}$. 
This minimax procedure therefore shows that $(\varphi,A)_{sph}=
(\phi_{-},0)$ is a sphaleron of the Abelian Higgs model.
To construct the NCL for the kink--type sphaleron $(\varphi,A)_{sph}=
(\phi_{+},0)$, one simply has to use 
\begin{equation}
\varphi^{-\infty}:\tau\mapsto e^{2i\tau}
\label{ac6}
\end{equation}
as the topologically nontrivial mapping.

To calculate the increase of the Chern--Simons number 
along the NCL, one has to treat the loop parameter as a 
(Euclidean) time dependent quantity 
$\tau=\tau(t)$ with $\tau(t=-\infty)=0$, $\tau(t=\infty)=\pi$. Inserting
the loop ansatz (\ref{a5}) into (\ref{Ncstime}),
one can split the double integral into 
a space ``volume'' and a ``surface'' integral, resulting in
\begin{equation}
\Ncs(t_0) = \frac{2}{\pi}\left\{\int_{-\infty}^{\infty}
\Omega_0\big|^{t=t_0}_{t=-\infty}dx + 
\int_0^{t_0}\Omega_1\big|_{x=-\infty}^{x=+\infty}dt\right\}
=\frac{\tau(t_0)}{\pi}+\frac{1}{2\pi}\sin 2\tau(t_0).
\label{a8}
\end{equation}
As expected, one finds $\Ncs=\frac12$ when the loop reaches the sphaleron,
whereas the entire loop has $\Ncs=1$.

\subsection{Gauging and nonperiodic boundary conditions}

The vacuum convention chosen above can be changed by {\em gauging the 
geometrical loop}, 
\begin{equation}
\bphi\rightarrow \tphi=g\bphi,\qquad \bA=0\rightarrow \tA=
ig^*\partial_xg. 
\label{aa8}
\end{equation}
One could, e.g., require $\tphi=g\bphi\rightarrow 1$
for $x\rightarrow -\infty$ {\em and} $x \rightarrow +\infty$, choosing the 
gauge \cite{FH}
\begin{equation}
g=e^{-i\tau\Lambda(x)},\qquad\Lambda(x)=\frac{2}{\pi}\arctan
\left(\frac{x}{\alpha}\right)+1
\label{ac8}
\end{equation}
($\alpha\neq 0$ arbitrary)
which yields $A(x)=\frac{2\tau}{\pi}
\frac{\alpha}{\alpha^2+x^2}$. Concerning its $\tau$--dependence, $g$ can be   
considered either as a set of static transformations parametrised by $\tau$ 
and applied to each corresponding static configuration along the loop
separately, or as one time dependent gauge transformation applied to the 
$\tau(t)$ time dependent loop as a whole. In the latter case, the temporal
gauge condition is violated, $A_0\equiv 0 \rightarrow \dot{\tau}\Lambda$.

In both cases, one finds for the gauge transformed Chern--Simons form
$\Omega_0\rightarrow\Omega_0+\frac14\tau\Lambda'$, 
$\Omega_1\rightarrow\Omega_1-\frac14\dot{\tau}\Lambda$. Inserting this into 
eq.\ (\ref{a8}), we find explicitly that the Chern--Simons number is gauge
invariant, i.e.\ also the gauged Manton loop $(\tphi,\tA)$
starts at a vacuum with
$\Ncs(\tau=0)=0$, reaches the sphaleron at 
$\Ncs\left(\tau=\frac{\pi}{2}\right)=\frac12$ and ends a vacuum
with $\Ncs(\tau=\pi)=1$. This is not surprising, as $\varrho=
\partial_{\mu}\Omega_{\mu}$ is a gauge invariant quantity.

The main new feature of the gauged geometrical 
loop is the representation of the
vacuum states between which the loop interpolates. Generalizing the gauged
geometrical loop $(\tphi,\tA)$ to a larger range of the loop
parameter $\tau\in\R$, the loop reaches a vacuum state
for any $\tau=n\pi$, $n\in\Z$, now given by
\begin{equation}
\tphi^{(n)}=\left(-\frac{\alpha-ix}{\alpha+ix}\right)^n   
\qquad \tA^{(n)}=\frac{n\alpha}{\alpha^2+x^2}
\label{a9}
\end{equation}
These vacua can be labeled by a ``topological charge'' defined as
\begin{equation}
Q:=\frac{2}{\pi}\int\Omega_0dx \quad \Rightarrow \quad
Q\big|_{(\tphi^{(n)},\tA^{(n)})}=\frac{1}{2\pi}\int \tA^{(n)} dx = n.
\label{a10}
\end{equation}
In the literature \cite{BS,C} this is usually the only ``topological number''
discussed in the context of the $SO(2)$ Abelian Higgs sphaleron. It
should be stressed that it is different from the Chern--Simons number
which is the proper quantity to be ivestigated in the context of the
sphaleron NCL and its relation to the instanton. The topological
charge used above to label the vacua makes use of the gauge--variant
part of the Chern--Simons form. Only in even dimensions does the Chern--Simons
form exhibit a gauge--variant part~\cite{desc}.
In these dimensions one can distinguish between
``small'' and ``large'' gauge transformations, depending on whether
they leave the ``topological charge'' (defined as space volume integral
of the zero component of the Chern--Simons form) invariant or not.

Another confusion about the $SO(2)$ Abelian Higgs found in the 
literature \cite{BS,C} concerns the use of periodic boundary conditions in the
space coordinates for the gauge fields, i.e.\ ``putting the fields on the 
circle'' instead of using the non--periodic vacuum structure (\ref{a9})
\cite{FH}. The sphaleron, however, is definitely
never a periodic object in space, but always the $\varphi^4$ kink, a 
contradiction which cannot be solved by simpy taking the limit of infinite 
period. 
 
The periodic solutions of $\varphi^4$ theory \cite{ManSam} and 
Goldstone theory \cite{BrihTom} in one dimension can be more gainfully
interpreted as the {\em periodic instantons in Euclidean quantum mechanics}.
These 
periodic solutions describe tunneling from thermally excited states, the 
temperature being given by the inverse period. At some value of the period, 
the periodic instantons reduce to the (time independent) sphaleron which in 
the case of Euclidean quantum mechanics is just a constant solution. This 
effect describes the phase transition between classical and quantum behaviour 
which is presently under intense investigation \cite{perinst}. Since periodic 
instantons (periodic in the Euclidean time, not in the spatial coordinate) 
are also known to exist in the Abelian Higgs model \cite{M}, one should be 
even more careful about the notion of periodicity in this model.

\subsection{AKY boundary loop construction}

A different technique for the construction of a NCL for the Abelian Higgs
sphaleron is 
motivated by the {\em ``static minimal energy path''} construction by
Akiba, Kikuchi and Yanagida (AKY) \cite{AKY} in the Weinberg--Salam theory.
This NCL is constructed by minimizing a general spherically symmetric static
ansatz with parameter-dependent boundary conditions, i.e.\ not the loop
ansatz containing 
the loop parameter, but the boundary conditions. We shall refer to 
this type of construction as ``boundary loop''.

The spherically symmetric AKY ansatz in one space dimension (in temporal 
gauge) is simple and only puts some restriction on the symmetry 
properties of the parameter functions,
\begin{equation}
\aphi(x)=H(|x|)+i\hat{x}K(|x|), \qquad \aA(x)=f(x)
\label{a11}
\end{equation}
with $\hat{x}=\sgn(x)$, i.e.\ $H(x)=H(|x|)$ is an even, $H(x)=\hat{x}H(|x|)$
an odd function of $x\in\R$. Under a gauge transformation $g=e^{i\Lambda(x)}$,
$H$ and $K$ are rotated, 
\begin{equation}
H\rightarrow H\cos\Lambda-K\sin\Lambda, \qquad
K\rightarrow K\cos\Lambda+H\sin\Lambda, 
\label{aa11}
\end{equation}
while $f\rightarrow f-\Lambda'$,
i.e.\ one can gauge $f=0$ without loss of generality in the ansatz (\ref{a11}).

Inserting the ansatz into the static energy functional yields
\begin{equation}
\CE[\aphi,\aA]=
\tilde{\CE}[H,K]=
\int\left[\frac12(H'+K')^2+\frac{1}{32}(1-H^2-K^2)^2\right]dx,
\label{a12}
\end{equation}
i.e. finite energy requires 
\begin{equation}
H(x\rightarrow\infty)=\cos q, \qquad K(x\rightarrow
\infty)=\sin q. 
\label{a12aa}
\end{equation}
Extrema of $\tilde{\CE}[H,K]$ with these boundary conditions
(and $H$ even, $K$ odd) are found only for $q=0$ or $q=\pi$, yielding the 
vacua $H=\pm 1$, $K=0$, and for $q=\frac{\pi}{2}$ with $H=0$,
$K(x)=\phi_+(x)=\tanh\left(\frac{x}{4}\right)$ which is the sphaleron
configuration.

The idea now is to construct a loop by increasing the boundary 
condition parameter $q(t)$
as a time--dependent quantity from $q(t=-\infty)=0$ to $q(t=\infty)=\pi$.
Inserting the ansatz in the Chern--Simons functional (\ref{a8}) and taking 
care of the boundary conditions for $H$, $K$ when evaluating 
$\Omega_1\big|_{x=-\infty}^{x=+\infty}$ yields
\begin{equation}
\Ncs=\frac{1}{2\pi}\int(KH'-HK')dx+\frac{q}{\pi},
\label{aa12}
\end{equation}
i.e.\ the
sphaleron has $\Ncs=\frac12$. One can minimise the static energy functional 
(\ref{a12}) for fixed Chern--Simons number by adding the Chern--Simons
functional with a Lagrange multiplier $\xi$ to the static energy functional,
leading to the Euler--Lagrange equations
\begin{equation}
H''+\xi K' + \frac18H(1-H^2-K^2) = 0, \qquad  
K''-\xi H' + \frac18H(1-H^2-K^2) = 0
\label{a13}
\end{equation}
which are solved with the above boundary conditions by
\begin{equation}
H(x)=\cos q,\qquad K(x)=\sin q \tanh\left(\frac{\sin q}{4}x\right),
\qquad \xi=8\cos q.
\label{aa13}
\end{equation} 
The energy
along this loop is found to be 
\begin{equation}
\tilde{\CE}(q)=\frac13\sin^3 q, 
\label{ab13}
\end{equation}
and the Chern--Simons number 
along the AKY loop for the Abelian Higgs model is 
\begin{equation}
\tilde{\CN}_{cs}(q)=\frac{q}{\pi}+\frac{1}{2\pi}\sin 2q.
\label{ac13}
\end{equation}
Therefore, the energy along the loop, $\tilde{\CE}(\tilde{\CN}_{cs})$ agrees
with the result (\ref{ab6},\ref{a8}) for the geometrical NCL. Hence there
is no advantage of lower energy along the boundary loop compared to
the geometrical loop, both provide equivalent 
``static minimal energy paths'' \cite{AKY}.

\section{$d=3$: Sphalerons in the Georgi--Glashow model}

The model is described by the  $SO(3)$ taking its values in the $SU(2)$
algebra with antihermitean generators $-\frac{i}{2}(\sigma_{\mu})=
-\frac{i}{2}\vec{\sigma}$ and a 
Higgs triplet field $(\phi^{\mu})=\vec{\phi}$ which we write in antihermitean
isovector representation, 
$\Phi=\vec{\phi}\cdot\left(-\frac{i}{2}\vec{\sigma}\right)$, in 
$d=3$ spacetime 
dimensions. The model is given by the Euclidean Lagrangian
\begin{eqnarray}
\CL & = &
\tr\left[-\frac14F_{\mu\nu}^2 - \frac12 (D_{\mu}\Phi)^2 + \frac{\lambda}{2}
\left(\Phi^2+\frac{\eta^2}{2}\right)^2\right] 
\label{b1} \\
& = & \frac12\left[\frac14\vec{F}^2_{\mu\nu}+
\frac12\left(D_{\mu}\vec{\phi}\right)^2+\frac{\lambda}{8}\left(\vec{\phi}^2-
2\eta^2\right)^2\right]
\label{b1pol}
\end{eqnarray}
with $F_{\mu\nu}=\partial_{[\mu}A_{\nu]}+[A_{\mu},A_{\nu}]
=\vec{F}_{\mu\nu}\cdot\left(-\frac{i}{2}\vec{\sigma}\right)$ and
$D_{\mu}\Phi=\partial_{\mu}\Phi+[A_\mu,\Phi]\Rightarrow
D_{\mu}\vec{\phi}=\partial_{\mu}\vec{\phi}+\vec{\phi}\wedge\vec{A}_{\mu}$,
with $\mu=1,2,3$.

>From the second form of the Lagrangian \cite{dilute} it is easy to identify
the particle spectrum of the theory which consists of photons, heavy 
charged bosons with mass $m_W=\sqrt{2}\eta$, and scalar neutral Higgs 
particles with mass $m_H=\sqrt{2\lambda}\eta$.

The model was previously
exploited by Polyakov \cite{dilute} to calculate quark confinement effects,
using a dilute gas of instantons. The instanton of the model is the
finite size spherically symmetric 
't Hooft--Polyakov monopole 
\cite{tHP} in three dimensions which is
characterised by integer Chern--Simons number
\begin{equation}
\Ncs=
\frac{1}{\sqrt{2}\pi\eta}\int\partial_{\mu}\Omega_{\mu}d^3x\quad \mbox{with}
\quad
\Omega_{\rho}=\frac14\epsilon_{\rho\mu\nu}\tr\left[\Phi F_{\mu\nu}\right]. 
\label{b1a}
\end{equation}
As the model is in an odd spacetime dimension, the Chern--Simons form 
$\Omega_{\mu}$ in this model, 
which is constructed from the Bogomol'nyi inequality 
\begin{equation}
\tr\left[F_{\mu\nu}-\epsilon_{\mu\nu\rho}D_{\rho}\Phi\right]^2>0\Rightarrow 
\CL>\partial_{\mu}\Omega_{\mu} \ ,
\label{b1b}
\end{equation}
has no gauge variant part \cite{desc}.

To discuss the sphalerons of the model \cite{Copeland}, 
we reduce the Langrangian (\ref{b1})
to the static Hamiltonian
\begin{equation}
\CH=\tr\left[-\frac14F_{ij}^2 - \frac12 (D_{i}\Phi)^2 +\lambda
\left(\Phi^2+\frac{\eta^2}{2}\right)^2\right] 
\label{b2}
\end{equation}
with $i,j,\ldots=1,2$.

\subsection{Sphaleron and boundary loop construction}
\label{so3aky}

Motivated by the AKY technique in the Weinberg--Salam model \cite{AKY},
we try the general static spherically symmetric ansatz for the Higgs field
to find the sphaleron and a corresponding NCL simultanously,
\begin{equation}
\Phi=i\frac{\eta}{\sqrt{2}}\left[H(r)\sigma_3+K(r)\hat{x}_i\sigma_i\right] \ ,
\label{b3}
\end{equation}
with $r^2=x^2_1+x^2_2$.

The $SO(2)$ gauge transformation $\Phi\rightarrow g^{-1}\Phi g$ with
\begin{equation}
g_{\Lambda}
=\exp\left\{\frac{i}{2}\epsilon_{ij} \hat{x}_i\sigma_j\Lambda(r)\right\}
\label{gGauge}
\end{equation}
rotates the parameter functions, 
\begin{equation}
H\rightarrow H\cos\Lambda-K\sin\Lambda,\qquad
K\rightarrow K\cos\Lambda+H\sin\Lambda. 
\label{b3a}
\end{equation}
This motivates the choice of a spatial spherically symmetric ansatz 
\begin{equation}
A_i=\frac{f_A+1}{r}\epsilon_{ik}\hat{x}_k\left(-\frac{i}{2}\sigma_3\right) +
\frac{f_B}{r}\epsilon_{ik}\hat{x}_k\left(-\frac{i}{2}\hat{x}_i\sigma_i\right)+
\frac{f_C}{r}\hat{x}_i \left(-\frac{i}{2}\epsilon_{kl}\hat{x}_k\sigma_l\right)
\label{b4}
\end{equation}
for the gauge field 
which transforms to an ansatz of the same type under 
$A_i\rightarrow g^{-1}A_ig+g^{-1}\partial_{\mu}g$, in particular,
\begin{equation}
f_A\rightarrow f_A\cos\Lambda-f_B\sin\Lambda,\quad
f_B\rightarrow f_B\cos\Lambda+f_A\sin\Lambda,\quad
f_C\rightarrow f_C-r\Lambda'. 
\label{b4a}
\end{equation}
The requirement of regularity at the origin implies 
\begin{equation}
f_A(0)=-1,\quad f_B(0)=0, \quad f_C(0)=0, \quad  H(0)=const, \quad K(0)=0,
\label{zero}
\end{equation}
Also the gauge 
transformation has to be regular at the origin, $\Lambda(0)=0$. 

In this gauge, inserting the ansatz (\ref{b3},\ref{b4}) reduces 
the static Hamiltonian to the following radial subsytem Hamiltonian:
\begin{eqnarray}
\tilde{H}_0& = & 
2\pi\Bigg\{
\frac{1}{4r}\left[\left(f_A^{\prime}-\frac{f_Bf_C}{r}\right)^2+
\left(f_B^{\prime}+\frac{f_Af_C}{r}\right)\right] 
\nonumber \\
& & {} +\frac{\eta^2}{2}\left[r\left(H^{\prime}-\frac{Kf_C}{r}\right)
+r\left(K^{\prime}+\frac{Hf_C}{r}\right)+\frac{1}{r}(Kf_A-Hf_B)^2\right]
\nonumber \\
& & {} +\frac{\lambda\eta^4}{4} r(1-H^2-K^2)^2\Bigg\}
\label{b5general}
\end{eqnarray}
The Euler--Lagrange equations are
\begin{eqnarray}
\left[r\left(H^{\prime}-\frac{Kf_C}{r}\right)\right]^{\prime} 
& = &
f_C\left(K^{\prime}+\frac{Hf_C}{r}\right)
-\frac{1}{r}f_B(Kf_A-Hf_B)
-\lambda\eta^2rH(1-H^2-K^2) 
\label{Eins}
\\
\left[r\left(K^{\prime}+\frac{Hf_C}{r}\right)\right]^{\prime} 
& = &
-f_C\left(H^{\prime}-\frac{Kf_C}{r}\right)
+\frac{1}{r}f_A(Kf_A-Hf_B)
-\lambda\eta^2rK(1-H^2-K^2) \nonumber \\
\label{Zwei}
\\
\left[\frac{1}{r}\left(f_A^{\prime}-\frac{f_Bf_C}{r}\right)\right]^{\prime} 
& = &
\frac{1}{r^2}f_C\left(f_B^{\prime}+\frac{f_Af_C}{r}\right)
+\frac{2\eta^2}{r}K(Kf_A-Hf_B)
\label{Drei}
\\
\left[\frac{1}{r}\left(f_B^{\prime}+\frac{f_Bf_C}{r}\right)\right]^{\prime} 
& = &
-\frac{1}{r^2}f_C\left(f_A^{\prime}-\frac{f_Bf_C}{r}\right)
-\frac{2\eta^2}{r}H(Kf_A-Hf_B)
\label{Vier}
\end{eqnarray}
and
\begin{equation}
0  = f_A\left(f_B^{\prime}+\frac{f_Bf_C}{r}\right)
-f_B\left(f_A^{\prime}-\frac{f_Bf_C}{r}\right)
 +2\eta^2r^2\left[H\left(K^{\prime}+\frac{Hf_C}{r}\right)
-K\left(H^{\prime}-\frac{Kf_C}{r}\right)\right]
\label{Fuenf}
\end{equation}
Note that the ansatz (\ref{b3},\ref{b4}) being radially symmetric, eqs.\ 
(\ref{Eins}--\ref{Fuenf}) are guaranteed 
consistent with the equations which are obtained by varying the original 
Hamiltonian (\ref{b2}) and inserting the ansatz (\ref{b3},\ref{b4}) afterwards.
This is in contrast with the situation in the Weinberg--Salam model where the
corresponding ansatz is not spherically symmetric and hence the 
consistency of the ansatz there must be checked. 

As the fifth equation (\ref{Fuenf}) is first--order, it is obviously not a 
dynamical equation, but a constraint on the system which is related to
a gauge transformation in the space of the parameter functions 
$(H,K,f_A,f_B,f_C)$ \cite{constraint}. 
This shows that this set of parameter functions has one 
redundant degree of freedom which is already clear from the fact that we 
kept the gauge freedom (\ref{gGauge},\ref{b3a},\ref{b4a}) in deriving 
the reduced Hamiltonian (\ref{b5general}).
Indeed, the gauge transformation generated by eq.\ (\ref{Fuenf}) is equivalent 
to the transformations (\ref{b3a},\ref{b4a}) derived from the 
$SO(3)$ gauge transformation $g$ of the original fields $(\Phi,A_{i})$
in (\ref{gGauge}). 

The gauge transformation (\ref{b3a},\ref{b4a}) can now be used to fix
$f_C\equiv 0$ without loss of generality in the ansatz (\ref{b3}). This
choice corresponds to ``radial gauge'' $\hat{x}_iA_i=0$. In this gauge,
the Hamiltonian (\ref{b5general}) simplifies to 
\begin{equation}
H_0=2\pi
\left\{\frac{1}{4r}(f_A^{\prime 2}+f_B^{\prime 2}) +
\frac{\eta^2}{2}\left[r(H^{\prime 2}+K^{\prime 2})+\frac{1}{r}(Kf_A-Hf_B)^2
\right]
+\frac{\lambda\eta^4}{4}r(1-H^2-K^2)^2\right\}
\label{b5}
\end{equation}

For $f_B\equiv 0$ and $H\equiv 0$ this reduces to the radial subsystem 
Euclidean Lagrangian of the Abelian Higgs model in two spacetime dimensions 
(\ref{abhiggsvortex}) supporting the (topologically stable) 
Abelian Higgs vortex. The embedding of this two dimensional 
$SO(2)$--Higgs vortex into the $SO(3)$--Higgs theory as a 
static object in three spactime dimensions, $(\Phi,A_i)_{sph}$ 
yields the sphaleron of the three dimensional model.

It is interesting to remark that the $SO(3)$ gauge field of this sphaleron 
tends to one half times a pure gauge at spatial infinity, $r\rightarrow\infty$:
\begin{equation}
(A_i)_{sph} \sim 
\frac{1}{r}\epsilon_{ik}\hat{x}_k\left(-\frac{i}{2}\sigma_3\right)
=\frac12 g_{\pi}^{-1}\partial_ig_{\pi}, \qquad 
g_{\pi}=\exp\left\{\frac{i}{2}\epsilon_{ij} \hat{x}_i\sigma_j\pi\right\},
\label{halfpuregauge}
\end{equation}
a property which the $SO(3)$ Higgs sphalerons shares with the instanton
of the same model \cite{dilute}. This is not surprising if we consider
that sphalerons and instantons are related objects.

Requiring finite energy 
\begin{equation}
\CE=\int\CH d^2x=\int H_0 dr<\infty
\label{energy}
\end{equation}
fixes the behaviour of the remaining parameter functions $(H,K,f_A,f_B)$ 
at spatial infinity to
\begin{eqnarray}
H(r\rightarrow\infty)=\cos q & & f_A(r\rightarrow\infty)=-\alpha(q)\cos q
\nonumber \\ 
K(r\rightarrow\infty)=\sin q & & 
f_B(r\rightarrow\infty)=-\alpha(q)\sin q 
\label{infty}
\end{eqnarray}
with $q\in[0,\pi]$.

The relation between $q$ and $\alpha=\alpha(q)$ 
can be determined from a more careful
analysis of the asymptotic behaviour of the parameter functions 
$(H,K,f_A,f_B)$, using the Euler--Lagrange equations of $H_0$ which are
\begin{eqnarray}
\left(\frac{f_A^{\prime}}{r}\right)^{\prime} 
& = & \frac{2\eta^2}{r}K(Kf_A-Hf_B)
\label{eins}
\\
\left(\frac{f_B^{\prime}}{r}\right)^{\prime} 
& = & -\frac{2\eta^2}{r}H(Kf_A-Hf_B)
\label{zwei}
\\
(r H^{\prime})^{\prime}
& = & -\frac{f_B}{r}f_B(Kf_A-Hf_B)-\lambda\eta^2 r H(1-H^2-K^2)
\label{drei}
\\
(r K^{\prime})^{\prime}
& = & \frac{f_A}{r}(Kf_A-Hf_B)-\lambda\eta^2 r K(1-H^2-K^2)
\label{vier}
\end{eqnarray}
Of course, these equations agree with (\ref{Eins}--\ref{Vier}), setting
$f_C=0$. The fifth equation (\ref{Fuenf}) reduces to
\begin{equation}
f_A\left(\frac{f_B^{\prime}}{r}\right)-
f_B\left(\frac{f_A^{\prime}}{r}\right)+
2\eta^2\left[H(r K^{\prime})-K(r H^{\prime})\right] = 0
\label{fuenf}
\end{equation}
which turns out to be an integration constant of the other four equations
(\ref{eins}--\ref{vier}).

The asymptotic analysis then fixes $\alpha(q)=\cos q$,
and one obtains the following asymptotic behaviour in the region $m_W r \gg1$: 
\begin{eqnarray}
H(r) & \sim & +\cos q - d_H\cos q\cdot (m_Wr)^{-\frac12}
e^{-m_H r} - d_W \cos q \sin q \cdot (m_Wr)^{-\frac32}e^{-m_W r}
\label{asymptH} \\
K(r) & \sim & +\sin q - d_H \sin q \cdot (m_Wr)^{-\frac12}
e^{-m_H r}+ d_W \cos^2 q \sin q \cdot (m_Wr)^{-\frac32}
e^{-m_W r}
\label{asymptK} \\
f_A(r) & \sim & - \cos^2 q + D_W\sin q \cdot (m_Wr)^{\frac12}e^{-m_W r}
\label{asymptfa} \\
f_B(r) & \sim & - \cos q \sin q - 
D_W\cos q \cdot (m_Wr)^{\frac12}e^{-m_W r}
\label{asymptfb} 
\end{eqnarray}
where $D_W$, $d_H$, $d_W$ are constants which can be determined numerically
by solving eqs.\ (\ref{eins}--\ref{vier}). 
It is easy to check that $m_W r =:\rho$ is the convenient dimensionless radial
variable in this model which is also used in all numerical calculations.

Solving eqs.\ (\ref{eins}--\ref{vier}) in the $m_W r\ll 1$ region yields
\begin{eqnarray}
H(r) & \sim & c_H+\frac14c_H(c_H^2-1)\cdot (m_Hr)^2 \label{smallH}\\
K(r) & \sim & c_K\cdot (m_Wr) \label{smallK} \\
f_A(r) & \sim & -1 + c_A \cdot (m_Wr)^2 \label{smallfA} \\
f_B(r) & \sim & c_B \cdot (m_Wr)^3 \label{smallfB} 
\end{eqnarray}
with $c_A$, $c_B$, $c_K$ constants which again have to be determined from
the numerical integration of the 
equations of motion, and $c_H=3\frac{c_B}{c_K}$.

Solutions of eqs.\ (\ref{eins}--\ref{vier}) can only be found for three 
particular values of $q$, $q\in [0,\pi]$. 
First, $q=0$ and $q=\pi$ allow the {\em vacuum configurations}
$H\equiv\pm 1$, $K\equiv 0$, $f_A\equiv -1$, $f_B\equiv 0$ with zero energy,
resulting in the vacuum fields
\begin{equation}
(\Phi,A_i)_{\pm}=\left(\pm i\frac{\eta}{\sqrt{2}}\sigma_3,0\right)
\label{vacua}
\end{equation}
Second for
$q=\frac{\pi}{2}$ the behaviour of the parameter functions at infinity 
(\ref{asymptH}--\ref{asymptfb}) and at the origin (\ref{smallH}--\ref{smallfB})
allow to set $f_B\equiv 0$ and $H\equiv 0$ such that the $SO(2)$ symmetric
Abelian Higgs vortex configuration is an element of the subspace of $SO(3)$
Higgs configurations given by the ansatz (\ref{b3},\ref{b4}) and the boundary 
conditions (\ref{zero},\ref{infty}) in the radial $f_C\equiv0$ gauge. 
This is the sphaleron solution.

To construct a loop from the subspace of $SO(3)$ Higgs configurations 
discussed above, we have to specify a one parameter subset 
$(\aPhi,\aA_i)_q$ parametrised by the boundary constant $q$. 
Three points of the loop are already fixed: The initial and final
vacua $(\aPhi,\aA_i)_{q=0,\pi}=(\Phi,A_i)_{\pm}$ and the sphaleron
$(\aPhi,\aA_i)_{q=\frac{\pi}{2}}=(\Phi,A_i)_{sph}$ given by the parameter 
functions $(f_A,K)_{sph}$. We now have to fix the parameter functions 
$(H,K,f_A,f_B)$ for all remaining values of $q$ such that $q$ becomes the
loop parameter. 

Instead of $q$, it is convenient to choose the Chern--Simons
number as parameter along the loop. Therefore, we next
calculate the Chern-Simons functional for
the set of configurations (\ref{b3},\ref{b4}) with boundary behaviour
(\ref{asymptH}--\ref{asymptfb}) at infinity and (\ref{smallH}--\ref{smallfB})
at the origin, treating $q=q(t)$ as time dependent parameter such that a loop
parametrised by $q$ starts at an initial vacuum with $q(t=-\infty)=0$ and 
ends at a final vacuum with $q(t=+\infty)=\pi$, whereas the sphaleron is 
reached at some time $t_{sph}$, $q(t_{sph})=\frac{\pi}{2}$.

Inserting the ansatz (in radial gauge $f_C\equiv 0$) 
into the Chern-Simons number functional
\begin{equation}
\Ncs(t_0) = \frac{1}{\sqrt{2}\pi\eta}\left\{\int_{\R^2}
\Omega_0\big|^{t=t_0}_{t=-\infty}d^2x + \int_{-\infty}^{t_0} dt
\lim_{r\rightarrow\infty} \int_{S^1(r)}\Omega_idS_i\right\}
\label{bCS}
\end{equation} 
we obtain 
\begin{equation}
\Ncs(q)=-\frac12\int\left(Hf_A^{\prime}+Kf_B^{\prime}\right)dr
+\frac12(1-\cos q),
\label{so3ncs}
\end{equation}
with $q=q(t_0)$. This yields $\Ncs(q=0)=0$ and $\Ncs(q=\pi)=1$, hence 
the Chern--Simons number increases by unity between the initial and the final 
vacuum. For the sphaleron, the surface integral in (\ref{bCS}) yields
$\Ncs\left(q=\frac{\pi}{2}\right)=\frac12$.

To complete the loop construction, we minimise the static energy (\ref{energy})
for fixed value of the Chern--Simons number $\Ncs$. This is done by
adding the integrand in (\ref{so3ncs}),
\begin{equation}
N_0=\frac12\left(Hf_A^{\prime}+Kf_B^{\prime}\right),
\label{lm}
\end{equation}
multiplied by a Lagrange multiplier $\xi$, to the reduced Hamiltonian 
$H_0$ in (\ref{b5}), and minimising 
\begin{equation}
F_0:=H_0+\xi N_0. 
\label{withmult}
\end{equation}
The corresponding 
Euler--Lagrange equations are
\begin{eqnarray}
\left(\frac{f_A^{\prime}}{r}\right)^{\prime} 
& = & \frac{2\eta^2}{r}K(Kf_A-Hf_B) - \xi H^{\prime}
\label{leins}
\\
\left(\frac{f_B^{\prime}}{r}\right)^{\prime} 
& = & -\frac{2\eta^2}{r}H(Kf_A-Hf_B) - \xi K^{\prime}
\label{lzwei}
\\
(r H^{\prime})^{\prime}
& = & -\frac{f_B}{r}f(Kf_A-Hf_B)-\lambda\eta^2 r H(1-H^2-K^2)
+\frac{\xi}{2\eta^2}f_A^{\prime}
\label{ldrei}
\\
(r K^{\prime})^{\prime}
& = & \frac{f_A}{r}(Kf_A-Hf_B)-\lambda\eta^2 r K(1-H^2-K^2)
+\frac{\xi}{2\eta^2}f_B^{\prime}.
\label{lvier}
\end{eqnarray}
The Lagrange multiplier changes the behaviour of the solutions at infinity
to
\begin{eqnarray}
H(r) & \sim & +\cos q - d_H\cos q\cdot (m_Wr)^{-\frac12}
e^{-\sqrt{m_H^2-\frac{\xi^2}{m_W^2}} r} 
- \xi d_W \sin q \cdot (m_Wr)^{\frac12}
e^{-\sqrt{m_W^2-\frac{\xi^2}{m_W^2}} r}
\label{lasymptH} \\
K(r) & \sim & +\sin q - d_H \sin q \cdot (m_Wr)^{-\frac12}
e^{-\sqrt{m_H^2-\frac{\xi^2}{m_W^2}} r}+ 
\xi d_W \cos q  \cdot (m_Wr)^{\frac12}
e^{-\sqrt{m_W^2-\frac{\xi^2}{m_W^2}} r}
\label{lasymptK} \\
f_A(r) & \sim & - \alpha(q)\cos q + \xi D_H \cos q \cdot (m_Wr)^{\frac12}
e^{-\sqrt{m_H^2-\frac{\xi^2}{m_W^2}} r} + D_W \sin q \cdot (m_Wr)^{\frac12}
e^{-\sqrt{m_W^2-\frac{\xi^2}{m_W^2}} r} \nonumber 
\\ \label{lasymptfa} \\
f_B(r) & \sim & - \alpha(q) \sin q + \xi D_H \sin q \cdot (m_Wr)^{\frac12}
e^{-\sqrt{m_H^2-\frac{\xi^2}{m_W^2}} r} - D_W \cos q \cdot (m_Wr)^{\frac12}
e^{-\sqrt{m_W^2-\frac{\xi^2}{m_W^2}} r}, \nonumber 
\\ \label{lasymptfb} 
\end{eqnarray}
and requiring  exponential decay restricts the range of the Lagrange 
multiplier to 
\begin{equation}
|\xi|<\min\{m_W^2,m_Wm_H\}.
\label{lmu}
\end{equation}
The asymptotic analysis no longer forces $\alpha(q)=\cos q$, and 
$\alpha$ along the loop is known only for the vacua and the sphaleron,
$\alpha(0)=1$, $\alpha(\pi)=-1$, $\alpha\left(\frac{\pi}{2}\right)=0$.
The general function $\alpha(q)$ has to be determined from the 
behaviour of the functions $f_A$, $f_B$ at $r\rightarrow\infty$.

We first consider the sphaleron and its boundary loop for $\lambda=1$
which happens to be a particularly simple case in the sense that the 
energy along the loop can be calculated analytically. Choosing
\begin{equation}
f_B\equiv 0, \qquad H\equiv \frac{\xi}{2\eta^2}
\label{self1}
\end{equation}
simplifies the equations (\ref{leins}--\ref{lvier}) to
\begin{eqnarray}
\left(\frac{f_A^{\prime}}{r}\right)^{\prime} & = & \frac{2\eta^2}{r}K^2f_A
\label{reins} \\
K^{\prime} & = & -\frac{1}{r}Kf_A
\label{rzwei} \\
f_A^{\prime} & = & \eta^2 r \left(1-\frac{\xi^2}{4\eta^2}-K^2\right)
\label{rdrei} \\
(rK^{\prime})^{\prime} & = & \frac{f_A^2}{r}K-
\eta^2rK\left(1-\frac{\xi^2}{4\eta^2}-K^2\right)
\label{rvier}
\end{eqnarray}
while the boundary conditions (\ref{zero},\ref{infty}) 
for the remaining functions $f_A$, $K$ become
\begin{eqnarray}
K(0) = 0 \qquad & &  \quad K(\rightarrow\infty)=
\pm\sqrt{1-\frac{\xi^2}{4\eta^4}} \nonumber \\
f_A(0) = -1, \quad & & \quad f_A(r\rightarrow\infty)=0 
\label{selfbound}
\end{eqnarray}
as $\cos q=\frac{\xi}{2\eta^2}$ from the boundary condition on
$H\equiv \frac{\xi}{2\eta^2}$ at $r\rightarrow\infty$.

It is easy to check that eqs.\ (\ref{rzwei}) and (\ref{rdrei}) are first 
integrals of eqs.\ (\ref{reins}) and (\ref{rvier}), hence a consistent solution
of the system (\ref{reins}--\ref{rvier}) can be found by solving the
first--order system (\ref{rzwei},\ref{rdrei}).

Solutions of eqs.\ (\ref{rzwei},\ref{rdrei}) saturate the inequalities
\begin{eqnarray}
\frac{1}{4r}\left[f_A^{\prime} - 
\eta^2 r \left(1-\frac{\xi^2}{4\eta^2}-K^2\right)\right]^2 & \ge & 0
\nonumber \\
\frac{\eta^2}{2}r\left[K^{\prime}  -\frac{1}{r}Kf_A\right]^2 & \ge & 0.
\label{self2}
\end{eqnarray}
The cross terms of the squares yield a lower bound for the Hamiltonian
(\ref{b5}) for $\lambda=1$ and $f_B\equiv0$, $H\equiv\frac{\xi}{2\eta^2}$:
\begin{eqnarray}
H_0\big|_{\lambda=1,f_B\equiv0,H\equiv\frac{\xi}{2\eta^2}} & = & 
2\pi\left\{\frac{1}{4r}f_A^{\prime 2} +
\frac{\eta^2}{2}\left[rK^{\prime 2}+\frac{1}{r}(Kf_A)^2
\right]
+\frac{\eta^4}{4}r\left(1-\frac{\xi^2}{4\eta^4}-K^2\right)^2\right\}
\label{selfham} \\
& \ge & \pi\eta^2 \frac{d}{dr}
\left\{f_A\left(1-\frac{\xi^2}{4\eta^4}-K^2\right)\right\}
\label{self3}
\end{eqnarray}
The resulting Hamiltonian (\ref{selfham}) is exactly of the same form as
the one dimensional reduced Lagrangian (\ref{abhiggsvortex}) of the
Abelian Higgs model, which becomes obvious by replacing the functions
$(a,h)$ by $(f_A ,K)$ formally, and by replacing
$1$ by $1-\frac{\xi^2}{4\eta^4}$ in the potential. Solutions to
this system (which saturate the bounds (\ref{self2}) are known numerically.

Exploiting the boundary behaviour (\ref{selfbound}) of these solutions,
one can find their energy depending on the Lagrange multiplier
$\xi$ from the saturated lower bound (\ref{self3}),
\begin{equation}
\CE = \int H_0\big|_{\lambda=1,f_B\equiv0,H\equiv\frac{\xi}{2\eta^2}} dr
=\pi\eta^2\left(1-\frac{\xi^2}{4\eta^4}\right).
\label{self4}
\end{equation}
Also the integral in the Chern--Simons number (\ref{so3ncs}) depends only 
on the boundary values for this particular loop, and we obtain
\begin{equation}
\Ncs=\frac12\left(1-\frac{\xi}{2\eta^2}\right) \: ,
\label{self5}
\end{equation}
resulting in an analytic expression for the energy along the boundary loop,
\begin{equation}
\CE(\Ncs) = 4\pi\eta^2\Ncs\left(1-\Ncs\right)\: .
\label{self6}
\end{equation}
As a result the slopes of $\CE(\Ncs)$ at the beginning and the end of the loop are
\begin{equation}
\left.\frac{\partial\CE(\Ncs)}{\partial\Ncs}\right|_{\Ncs=0}=4\pi\eta^2,\qquad
\left.\frac{\partial\CE(\Ncs)}{\partial\Ncs}\right|_{\Ncs=1}=-4\pi\eta^2.
\label{slopes}
\end{equation}

For coupling constants $\lambda>1$, the full set of equations 
(\ref{leins}--\ref{lvier}) has to be be solved
numerically, using $\rho=\sqrt{2}\eta r$ as rescaled dimensionless radial 
variable and eliminating $q$ from the boudary conditions at infinity.
In practice we replaced the boundary conditions (\ref{infty}) 
which also involve
the unknown function $\alpha(q)$ by
\begin{eqnarray}
H^2+K^2 & \stackrel{r\rightarrow\infty}{\longrightarrow} & 1 \nonumber \\
Kf_A+Hf_B & \stackrel{r\rightarrow\infty}{\longrightarrow} & 0
\nonumber \\
Hf_A^{\prime}+Kf_B^{\prime} & \stackrel{r\rightarrow\infty}{\longrightarrow} &
0 \nonumber \\
f_A^{\prime} & \stackrel{r\rightarrow\infty}{\longrightarrow} &
0
\label{varinfty}
\end{eqnarray}
and solved eqs.\ (\ref{leins}--\ref{lvier}) for given value of $\xi$, 
identifying $q$ and $\alpha(q)$ from the asymptotic behaviour of the 
numerical solutions afterwards.
 
\begin{figure}
\begin{center}
\includegraphics[bb=2cm 8.5cm 19cm 23cm,angle=0]{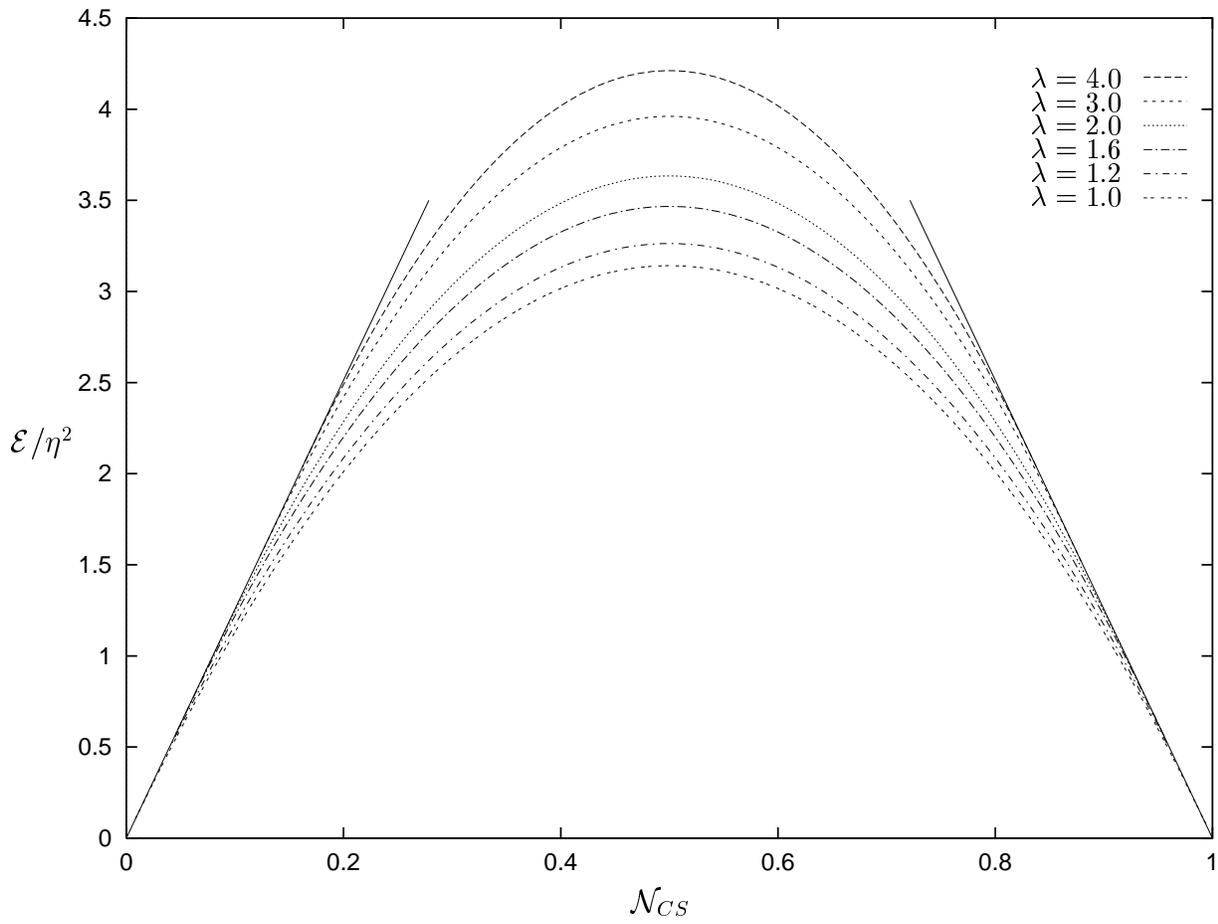}
\end{center}
\caption{The energy along the boundary loop, $\CE(\Ncs)$, for
$\lambda=1.0;1.2;1.6;2.0;3.0;4.0$.
The straight lines show the slopes of the energy curve at 
the vacua $\Ncs=0,1$.}
\label{loopfig1}
\end{figure}

The energy $\CE$ and
Chern--Simons number $\Ncs$ can be computed for these solutions. This finally
yields a ``static minimal energy loop'' connecting two topologically
neighbouring vacua through the sphaleron configuration which corresponds
to the vortex of the Abelian--Higgs model. The energy along the loop
which can be parametrised by the Chern--Simons number, $\CE(\Ncs)$, is
shown in Fig.\ \ref{loopfig1}, for several values of the ratio
$\frac{m_H^2}{m_W^2}=\lambda>1$. In this parameter region, the
Lagrange multipier is restricted to $|\xi|<\xi_c=2\eta^2$. 
$\CE(\Ncs)$ approaches the vacua $\CE=0$ for 
$\xi\rightarrow\mp 2\eta^2$ with slopes
\begin{equation}
\left.\frac{\partial\CE(\Ncs)}{\partial\Ncs}\right|_{\Ncs=0;1}=\pm4\pi\eta^2
\label{sslop}
\end{equation}
independently of $\lambda$, because the slope of the energy curve approaching 
the vacua is given by $\pm 2\pi\xi_c$ \cite{AKY}.

The symmetry $\CE(\Ncs)=\CE(1-\Ncs)$ is related to the invariance
of $H_0$ and $F_0$ in eqs.\ (\ref{b5}.\ref{withmult}) under the transformation
\begin{eqnarray}
H \mapsto -H \qquad & & \qquad f_A \mapsto \quad\! f_A \nonumber \\
K \mapsto\quad K \qquad & & \qquad f_B \mapsto -f_B
\label{symme}
\end{eqnarray}
which requires
\begin{equation}
q\mapsto \pi-q,\qquad \alpha(\pi-q)=\alpha(q).
\label{symme1}
\end{equation}
It is easy to check that this transfrmation yields
\begin{equation}
\Ncs\mapsto 1-\Ncs
\label{symme2}
\end{equation}
while the energy is invariant, yielding the symmetry of the energy along the
boundary loop mentioned above.
 
The boundary loop of the $SO(3)$--Higgs model we constructed  
is different from the original AKY construction in the 
Weinberg--Salam model \cite{AKY}. The Weinberg--Salam sphaleron has a pure
gauge connection
at infinity. This allowed a boundary loop construction \cite{AKY}
where all gauge fields
along the loop tend to a pure gauge, in particular, gauge and Higgs fields at 
infinity could be constructed as gauge transformation of the initial
vacuum configuration, using the gauge transformation which rotates the
parameter fields and allows to gauge $f_C\equiv 0$. This gauge transformation
$g_{\Lambda}$ (\ref{gGauge}) also exists in the $SO(3)$--Higgs boundary loop 
construction, but it can not be used to construct the loop configurations at 
infinity from the initial vacuum. In fact, applying $g_{\Lambda=q}$ to
the initial vacuum $(\aPhi,\aA_i)_{q=0}=(\Phi,A_i)_+$ yields the 
correct behaviour of the Higgs field, but not of the gauge field along
the loop at spatial infinity,
\begin{eqnarray}
(\aPhi,\aA_i)_{q=0} & = &  \left(i\frac{\eta}{\sqrt{2}}\sigma_3,0\right)
\nonumber \\
& \stackrel{g_q}{\mapsto} & 
\left(i\frac{\eta}{\sqrt{2}}(\cos q \sigma_3 + \sin q \hat{x}_i \sigma_i),
\frac{1-\cos q}{r}\epsilon_{ik}\hat{x}_k\left(-\frac{i}{2}\sigma_3\right) 
-\frac{\sin q}{r}
\epsilon_{ik}\hat{x}_k\left(-\frac{i}{2}\hat{x}_i\sigma_i\right)\right) \ ,
\nonumber \\
\label{atinfinity}
\end{eqnarray}
as one can see by comparing with the Higgs and gauge fields along the loop at
spatial infinity, i.~e.
\begin{eqnarray}
(\aPhi,\aA_i)_q \quad & \vspace{-1cm} \stackrel{r\rightarrow\infty}{\sim} &
\nonumber \\
& &  \hspace*{-1.5cm} 
\left(i\frac{\eta}{\sqrt{2}}(\cos q \sigma_3 + \sin q \hat{x}_i \sigma_i),
\frac{1-\alpha(q)\cos q}{r}
\epsilon_{ik}\hat{x}_k\left(-\frac{i}{2}\sigma_3\right) 
-\frac{\alpha(q) \sin q}{r}
\epsilon_{ik}\hat{x}_k\left(-\frac{i}{2}\hat{x}_i\sigma_i\right)\right) \ .
\nonumber \\
\label{gaugeloop}
\end{eqnarray}
Thi is due to the fact that the sphaleron, which has to be a loop
configuration, is one half times a pure gauge at infinity in accordance 
with the 
corresponding behaviour of the instanton.

We expect that this is a general feature of models which support 
both instanton and
sphaleron, i.e.\ the sphaleron behaves like the instanton at spatial infinity,
a fact which must be reflected in the boundary loop construction.

\subsection{General vorticity sphalerons and geometrical loop construction}
\label{so3manton}

In general, the Abelian Higgs vortex has vorticity or winding number
$N\in \N$. So far, we considered only $N=1$ for the $SO(3)$ sphaleron, but
the generalisation to sphalerons of arbitrary integer vorticity can 
easily be achieved. We use this generalisation in our presentation of 
the second, geometrical loop construction. 

The starting point of the geometrical loop construction is the sphaleron 
ansatz which we generalise to vorticity $N$, replacing the spatial unit 
vector $\hat{x}$ by $\hat{n}=(\cos N\phi,\sin N\phi)$ in the ansatz,
\begin{equation}
\Phi=i\frac{\eta}{\sqrt{2}} h(r) \hat{n}_i\sigma_i,\qquad
A_i=\frac{N}{r}f(r)\epsilon_{ik}\hat{x}_k\left(-\frac{i}{2}\sigma_3\right),
\label{b6}
\end{equation}
where we use $h(r)$ and $f(r)$ as parameter functions \cite{Manton}.
Inserting the ansatz into the Hamiltonian (\ref{b2}) yields the radial
subsystem Hamiltonian 
\begin{equation}
H_{sph}=
2\pi\left\{\frac14\frac{N^2}{r}f^{\prime 2}+\frac{\eta^2}{2}\left[
r h^{\prime 2}+\frac{N^2}{r}h^2(1-f)^2\right]+\frac{\lambda\eta^4}{4}
r(1-h^2)^2\right\}.
\label{b7}
\end{equation}
Finite energy and regularity at the origin require 
\begin{equation}
f(r\rightarrow\infty)=1,\quad h(r\rightarrow\infty)=1,\qquad
f(0)=0,\quad h(0)=0. 
\label{rand}
\end{equation}
and the solutions are equivalent to the vorticity $N$ Abelian Higgs vortices
as expected.

The geometrical loop construction starts from requiring finite energy, hence
the potential term in the Hamiltonian (\ref{b2}) forces  
$|\Phi|=\eta\Rightarrow \Phi\in S^2_{Higgs}$ at spatial infinity 
$(\R^2)^{\infty}=S^1_{space}$, described by the angular coordinate $\phi$.
Therefore we consider the Higgs field for $r\rightarrow\infty $ 
and add a single loop parameter $\tau\in S^1_{loop}$ to construct a nontrivial 
mapping, 
\begin{equation}
\Phi^{\infty}:S^1_{space}\times
S^1_{loop}\rightarrow S^2_{Higgs}. 
\label{b7a}
\end{equation}
The simplest choice for this mapping with vorticity $N$ is
\begin{equation}
\Phi^{\infty}=i\frac{\eta}{\sqrt{2}}\vec{p}\cdot\vec{\sigma}, \qquad
\vec{p}=\left(\begin{array}{c}
\sin\tau\cos N\phi\\ \sin^2\tau\sin N\phi + \cos^2 \tau \\
\sin\tau\cos\tau(\sin N\phi -1)
\end{array}\right)
\label{b8}
\end{equation}

This fixes also the gauge field at infinity since the covariant derivative
in the Hamiltonian (\ref{b2}) has to vanish, requiring
\begin{equation}
A_i^{\infty}:=A_i(r\rightarrow\infty)=-\frac{1}{2\eta^2}[\Phi^{\infty},
\partial_i\Phi^{\infty}].
\label{b8a}
\end{equation}
According to the results of  Section \ref{so3aky}, the gauge field at infinity
along the loop is not a pure gauge, in particular, it tends to half a pure
gauge for $r\rightarrow\infty$ for the sphaleron, $\tau=\frac{\pi}{2}$.

For $\tau=0$ and $\tau=\pi$, the loop has to start and end in the 
vacuum which for convenience \cite{Manton} we choose to be 
\begin{equation}
(\Phi,A_i)_{vac}=\left(i\frac{\eta}{\sqrt{2}}\sigma_2,0\right)
\label{b8b}
\end{equation}
in contrast to the choice of the vacua in Section \ref{so3aky}. This yields
the geometrical NCL ansatz
\begin{equation}
(\lPhi,\lA_i)_{\tau} = \left(i\frac{\eta}{\sqrt{2}}[(1-h(r))\vec{t}+h(r)
\vec{p}],-\frac{1}{2\eta^2}f(r)A_i^{\infty}\right) 
 \qquad {\rm with} \ \ 
\vec{t}=\left(\begin{array}{c}
0\\ \cos^2\tau\\
-\sin\tau\cos\tau
\end{array}\right)
\label{b9}
\end{equation}
The boundary conditions (\ref{rand}) then ensure 
\begin{equation}
(\lPhi,\lA_i)_{\tau}\stackrel{r\rightarrow\infty}{\longrightarrow}
(\Phi^{\infty},A_i^{\infty}) \ ,
\label{bbb9}
\end{equation}
and $\vec{t}$ is chosen such that 
\begin{equation}
(\lPhi,\lA_i)_{\tau=0}=(\lPhi,\lA_i)_{\tau=\pi}=(\lPhi,\lA_i)_{vac} \ .
\label{equation}
\end{equation}
One can also
check that $\Phi^2$ is a radial function of $r$ only which is necessary
for the consistency of the ansatz.

Inserting the loop ansatz (\ref{b9})
into the static Hamiltonian (\ref{b2}) yields
\begin{equation}
H_{\tau}=
2\pi\sin^3\tau\left\{\frac14\frac{N^2}{r}f^{\prime 2}+\frac{\eta^2}{2}\left[
r h^{\prime 2}+
\frac{N^2}{r}\left(h-f(h\sin^2\tau+\cos^2\tau)\right)^2
\right]+\frac{\lambda\eta^4}{4}\sin^2\tau r(1-h^2)^2\right\}.
\label{b10}
\end{equation}
For $\tau=\frac{\pi}{2}$, this geometrical 
loop ansatz reduces to the sphaleron 
ansatz (\ref{b6}), and of course $H_{\tau=\frac{\pi}{2}}=H_{sph}$, whereas
$H_{\tau=0}=0=H_{\tau=\pi}$ for the vacua at the beginning and the end of the
loop.

$H_{\tau}$ can be minimized numerically
for any value $\tau\in[0,\pi]$, yielding again a
``static minimal energy path'' connecting two neighbouring vacua through the 
sphaleron, the energy along the loop being $\CE(\tau)=\int H_{\tau}dr$.

One can also calculate the Chern--Simons number along the NCL, treating
$\tau=\tau(t)$ as time dependent with $\tau(-\infty)=0$, $\tau(\infty)=\pi$. 
For the Manton loop ansatz (\ref{b9}), both the volume and the surface 
integrals in eq.\ (\ref{bCS}) contribute, resulting in  
\begin{equation}
\Ncs(\tau) =\frac{N}{2}\cos\tau
\sin^2\tau\int f'(1-h) dr + \frac{N}{2}(1-\cos\tau). 
\label{b10a}
\end{equation}
For the sphaleron $\tau=\frac{\pi}{2}$, only the surface integral contributes 
to $\Ncs=\frac{N}{2}$, whereas the loop as a whole has $\Ncs=N$.
The vorticity $N$ sphaleron therefore is the saddle point of a loop
connecting to vacua with Chern--Simons number difference $N$.

\begin{figure}
\begin{center}
\includegraphics[bb=2cm 8.5cm 19cm 23cm,angle=0]{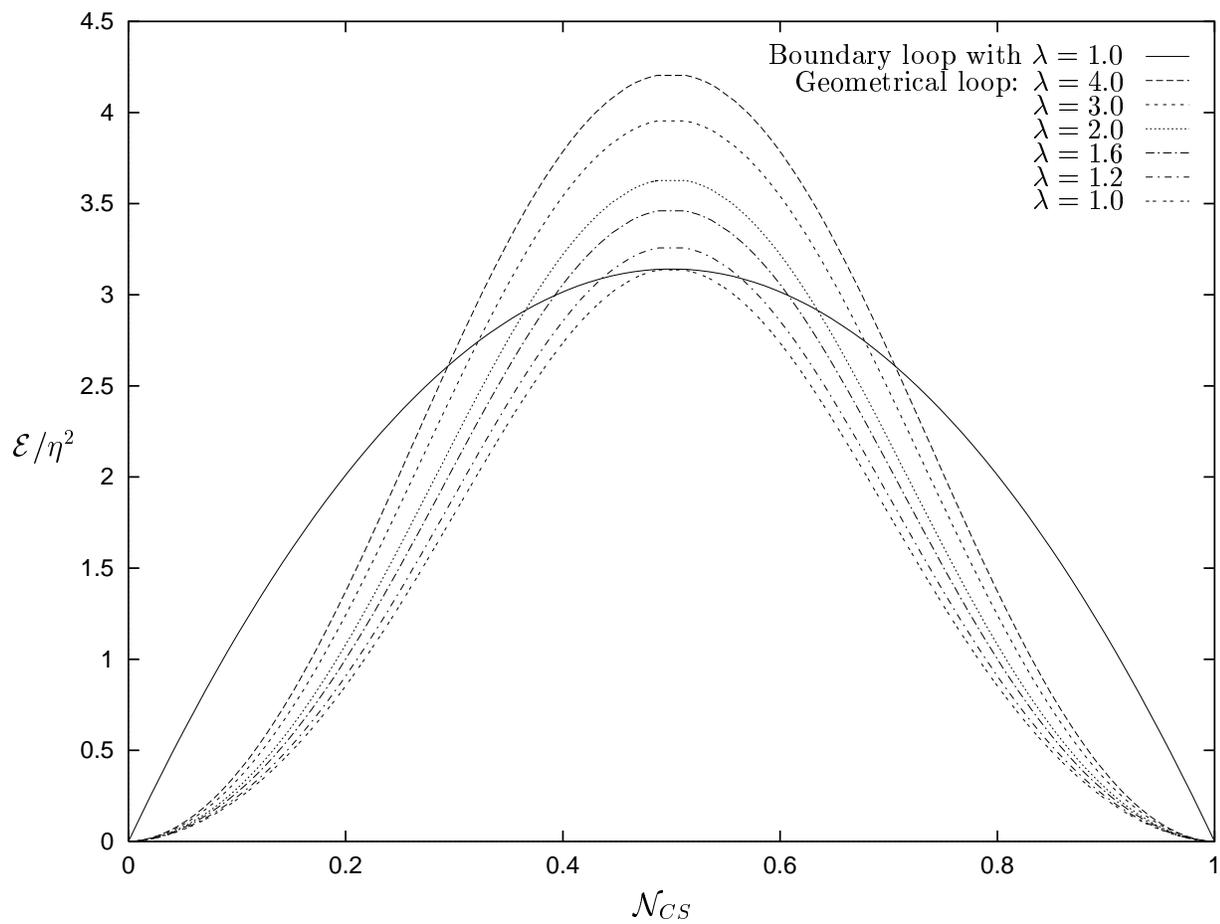}
\end{center}
\caption{The energy along the geoemtrical loop, $\CE(\Ncs)$, with vorticity
$N=1$ and $\lambda=1.0;1.2;1.6;2.0;3.0;4.0$, and the energy along the 
boundary loop for $N=1$, $\lambda=1.0$}
\label{loopfig2}
\end{figure}

The ``static minimal energy path'' of the $SO(3)$ gauge Higgs
sphaleron in the geometrical construction, $\CE(\Ncs)$, is shown in Fig.\
(\ref{loopfig2}) with vorticity $N=1$ and several values of $\lambda$. 
Comparing the $\lambda=1$ geometrical loop with the corresponding boundary 
loop, we find that the energy along the geometrical loop is lower than that
along the boundary loop except at the sphaleron and the vacua where they are
equal. This result holds also for $\lambda>1$.

\begin{figure}
\begin{center}
\includegraphics[bb=2cm 8.5cm 19cm 23cm,angle=0]{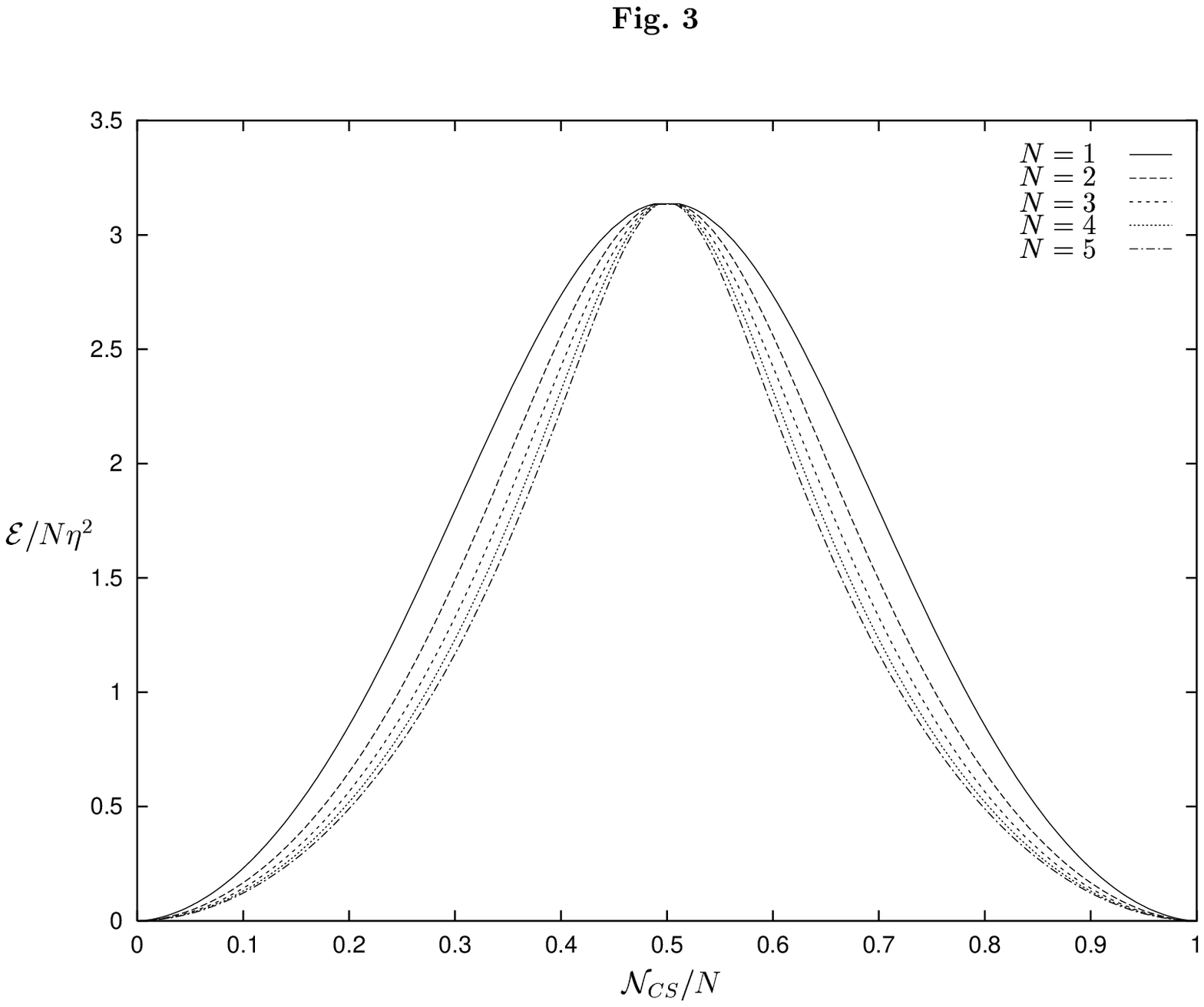}
\end{center}
\caption{The energy along the geometrical loop, 
$\frac{\CE\left(\frac{\Ncs}{N}\right)}{N}$, with
$\lambda=1.0$ and vorticity $N=1;2;3;4;5$}
\label{loopfig3}
\end{figure}

Fig.\ \ref{loopfig3} shows the geometrical loop for several values of the
vorticity.

\subsection{Magnetic and electric properties of Georgi--Glashow sphalerons}

Since the effective equations governing the sphaleron solution of the 3
dimensional Georgi--Glashow model are those of the Abelian Higgs vortex with
vorticity $N$ (hence also magnetic charge), we might
expect that the sphaleron itself may carry magnetic charge.

The electromagnetic field strength of the 3 dimensional Georgi--Glashow model
is the well-known 't Hooft electromagnetic tensor~\cite{tHP,Copeland}
\begin{equation}
{\cal F}_{\mu\nu}:=-\frac{\tr[\Phi F_{\mu\nu}]}{\sqrt{-\tr[\Phi^2]}} = 
\frac{\vec{\phi}\cdot\vec{F}_{\mu\nu}}{|\vec{\phi}|} \: .
\label{b11}
\end{equation}
Inserting the sphaleron configuration (\ref{b6}) into (\ref{b11}) yields
${\cal F}_{\mu\nu}\equiv 0$, therefore the sphaleron itself carries neither
magnetic nor electric charge. The situation is different if we consider
the complete NCL as the configuration dependent on the time parameter.
For example, along the boundary loop presented in Section \ref{so3aky} with
$q=q(t)$ we have nonvanishing electric and magnetic fields,
\begin{eqnarray}
B & = & {\cal F}_{12} = -\frac{1}{r}\frac{Hf_A^{\prime}+Kf_B^{\prime}}
{\sqrt{H^2+K^2}} \label{fmag} \\
E_i & = & {\cal F}_{0i} = -\epsilon_{ij}
\frac{\hat{x}_j}{r}\frac{H\dot{f}_A+K\dot{f}_B}
{\sqrt{H^2+K^2}}. \label{fel}
\end{eqnarray}
It follows, by symmetry, that the electric charge of the NCL,
\begin{equation}
Q_e := \lim_{r\rightarrow\infty} \int_{S^(r)} \vec{E}\cdot d\vec{S},
\label{b15}
\end{equation}
is always zero, whereas the NCL configurations may acquire magnetic
charge
\begin{equation}
Q_m := \int B d^2x
\label{b15aa}
\end{equation}
at some values of the time parameter. Clearly, the magnetic charge of the
sphaleron and the initial and final vacua vanish according to the
values that the functions $f_A ,f_B , K$ and $H$ take for these configurations.
The same conclusion is arrived at also by inspecting the magnetic flux of the
vorticity $N$ configuration calculated in terms of the geometric loop
parameter $\tau$
\be
\label{b15aaa}
Q_m = \sin^2 \tau \: \cos \tau \: N\: \int_0^{\infty}(1-h)f' dr \: ,
\ee
which vanishes at the sphaleron, $\tau =\frac{\pi}{2}$ and the vacuua
$\tau =0$ and $\tau =\pi$.

One might think at this point that there is no reason why there should be no
nonvanishing electric flux if the temporal gauge condition $A_0 =0$ were
relaxed in the spirit of the Julia-Zee dyon~\cite{JuliaZee}. In the
$A_0 \neq 0$ one can solve the full Euler-Lagrange equations in Minkowski
space, of
\begin{equation}
\CL_M=\tr\left[
-\frac14 F_{\mu\nu}F^{\mu\nu} + \frac12 D_{\mu}\Phi D^{\mu}\Phi -
\lambda\left(\phi^2+\frac{\eta^2}{2}\right)^2\right],
\label{b12}
\end{equation}
in the static limit using the radially symmetric restriction of the fields
according to the Ansatz
\begin{eqnarray}
\Phi & = & i \frac{\eta}{\sqrt{2}} h(r) \hat{x}_i\sigma_i \nonumber \\
A_0 & = & i \frac{\eta}{\sqrt{2}} g(r) \hat{x}_i\sigma_i, 
\qquad A_i=  \frac{f(r)}{r} \epsilon_{ij}\hat{x}_j 
\left(-\frac{i}{2}\sigma_3\right).
\label{b13}
\end{eqnarray}
To find solutions that lead to nonvanishing electric flux $Q_e$,
the solution must have the following asymptotic behaviour
\begin{equation}
g(r) \sim d_g + \sqrt{2}\pi\eta Q_e \ln r, \qquad m_W r \gg 1 \: ,
\label{b14}
\end{equation}
where $d_g$ is a constant. The asymptotic analysis of the equations, not
exibited here, indeed confirms the behaviour (\ref{b14}). These equations have been
integrated in \cite{Copeland} numerically.

However, this static electrically charged solution can not be called a 
sphaleron since the logarithmic behaviour of $g(r)$ 
(which is a consequence of $\ln r$ being the fundamental solution of the 
Laplace operator in two 
dimensions) destroys the integrability of the static Hamiltonian, i.e.\
this electrically charged classical configuration does not have finite
energy, rendering it useless as a  pseudoparticle in 
quantum field theory. Technically, this is related to the well--known fact
that the energy integral of point charges in two dimensional electrodynamics
is divergent.

\section{Summary and discussion}

We have analysed the first two in the hierarchy of 
$SO(d)$ gauged $d$ dimensional Higgs models
where the Higgs fields are in the $d$-dimensional 
vector representation of $SO(d)$. In $d=2$ and
$3$ these are the familiar Abelian Higgs model and 
the Georgi-Glashow model respectively. The
reason for having chosen these models is that they 
both support topologically stable finite
action solutions which we interpret as the instantons, 
and, they are expected to support
sphaleron solutions in the static limit, which we find 
indeed to be the case. Thus we have
analysed the first two Higgs models in this hierarchy, 
which support both instantons and
sphalerons. This is the main criterion of the work, and 
our motivations for it are explained in
the Introduction. Our analysis points the way to tackling 
the $d=4$ case~\cite{OT} which is of
some definite physical relevance inasfar as it promises a 
Coulomb instanton gas~\cite{so4inst},
but which is considerably more complex.

Since the instantons of these two models are well known 
solutions, namely the Nielsen--Oleson
vortices~\cite{NO} and the 't Hooft--Polyakov 
monopole~\cite{tHP}, the analysis of this paper
is restriced to the study of the sphalerons of these models.

Indeed, the sphalerons of the
Abelian Higgs model have been studied extensively in the 
literature~\cite{BS,C,FH}, but we give
our own version here. The reason for this is firstly so 
that the construction of the sphalerons
of both models should proceed in the same lines, 
especially since one of criteria is to map the
way to carry this analysis to the $d=4$ case. 
Secondly, our construction of the sphaleron of
the Abelian Higgs model follows exactly the same 
procedure as those used in the corresponding
work for the Weinberg-Salam model, namely that of 
constructing a NCL carried out by Manton
\cite{Manton}, and also that of constructing the 
path of finite energy configurations carried
out by Akiba {\it et.al.}~\cite{AKY}. This contrasts with 
the presentation in the literature~\cite{BS,C,FH}
where the procedure is quite different from the 
case of the Weinberg-Salam model~\cite{Manton,
Klinkhamer, AKY}, in particular imposing periodic 
boundary conditions~\cite{BS} unnecessarily.
This part of our work is presented in Section 2, 
where we have given both construction of the
sphaleron, namely that of Manton and 
Klinkhamer~\cite{Manton, Klinkhamer} and that of Akiba
{\it et.al.}~\cite{AKY}. In addition, we have made a contrast 
of our procedures with those existing in
the literature. Our procedure here making it possible 
to formulate this problem in complete
parallel to the Weinberg-Salam case is made possible by 
our use of the systematically
derived~\cite{desc} Chern-Pontryagin density used in 
calculating the Chern-Simons number.

In the larger part of this work we have presented the 
construction of the sphaleron in the
three dimensional Georgi-Glashow model. This is the work 
in Section 3. Again we have presented
the construction of the sphaleron in both the Manton and 
Klinkhamer~\cite{Manton, Klinkhamer}
procedure and the Akiba {\it et.al.}~\cite{AKY} one. It is very 
interesting that in the former case
\cite{Manton, Klinkhamer}, we find that the sphaleron is 
a solution to the effective Abelian
Higgs model in the two spatial dimensions, embedded in the 
$SO(3)$ model. In that case we
have constructed a family of sphalerons characterised by a 
vortex number $N$, availing of the
fact that radially symmetric fields in two dimensions have 
integer vorticity. We have also
inquired whether this sphaleron has magnetic flux related 
to its vorticity $N$ and have found
that the sphaleron itself, as well as the vacuua it falls 
between, have zero magnetic flux,
but that intermediate field configurations do have 
nonvanishing magnetic flux. We have also
verified that by relaxing the temporal gauge, one can find 
solutions to the static field
equations leading to a nonvanishing electric 
flux~\cite{Copeland}, but in that case the energy
diverges logarithmically and hence there is no sphaleron 
solution that can be gainfully employed
in semiclassical field theory. In addition to this, we 
have pursued the construction in the
procedure of Akiba {\it et.al.}~\cite{AKY} and have constructed 
the finite energy path interpolating
between the two vacuua with the sphaleron at the top of 
this path with Chern--Simons number equal
to $1\over 2$. An interesting circumstance here is that 
when the coupling constant of the Higgs
self-interaction potential in the Georgi-Glashow model 
takes the critical value for which the
(static) embedded Abelian Higgs model can saturate the 
corresponding Bogomol'nyi bound, the
finite energy path can be constructed analytically without 
recourse to numerical computations.

Our analysis of these two models has clarified the 
similarities and differences between the
sphalerons of the Weinberg-Salam model and those of the 
models in our hierarchy of $SO(d)$ Higgs
models. In this respect, the $d=3$ case is the most enlightening. We have
learnt that the structure of the geometric loop~\cite{Manton,Klinkhamer}
behaves very much like that of the Weinberg-Salam model, and that the
boundary loop~\cite{AKY} also behaves similarly inspite of the fact that the
actual boundary conditions are quite different in the two cases. In the process,
we have given a unified treatment for the construction of energy loops
interpolating between vacuua for all models, bringing the treatment of the Abelian
Higgs case into line with the others. This we have done employing both types of
energy loops, geometrical~\cite{Manton} and boundary~\cite{AKY}, whence we have
learnt that the geometric loop is lower than the boundary loop everywhere except
at the sphaleron and the vacuua where they are equal. Presumably this is the case
also for the Weinberg-Salam model.

As a final remark we emphasise the similarity between the boundary loop
construction for the Winberg-Salam and the Georgi-Glashow models --
in both of them the functions $f_B$ and $H$ are excited on the boundary loop.
In addition both theories tend to sigma models in the limit of the Higgs masses
becoming infinit. With these two similarities in place, it seems that there may
well be bisphalerons \cite{BK,J} in the static 3 dimensional 
Georgi-Glashow model.

\bigskip

\noindent
{\bf Acknowledgments}

\noindent
This work was carried out in part under 
Basic Science Research project SC/97/636 of
FORBAIRT. FZ acknowledges a Presidency of Ireland Postdoctoral
Fellowship from the HEA. We are very grateful to Yves Brihaye for
valuable discussions.

\end{document}